\documentclass[prb,showpacs,amsmath,amssymb,twocolumn,preprintnumbers]{revtex4}

\usepackage{graphicx}
\usepackage{dcolumn}
\usepackage{bm}

\preprint{submitted to Journal of Computational and Theoretical
Nanoscience for a special issue on Low-Dimensional Systems}

\begin{document}

\title{Effect of surfaces and interfaces on the electronic, magnetic and
gap-related properties of the half-metal Co$_2$MnSn}

\author{I. Galanakis}\email{galanakis@upatras.gr}
\affiliation{Department of Materials Science, School of Natural
Sciences, University of Patras,  GR-26504 Patra, Greece}

\date{\today}

\begin{abstract}
We present state-of-the-art electronic structure calculations for
the Co$_2$MnSn full-Heusler alloy. We show that in its bulk form
it is a half-metallic ferromagnet with the Fermi level being
located within a tiny gap of the minority-spin density of states.
Moreover the alloy shows the Slater-Pauling behavior with a total
spin magnetic moment in the unit cell of 5 $\mu_B$. In the case of
the (001) surfaces, the broken bonds at the surface form a
minority band pinned exactly at the Fermi level destroying the
half-metallicity. Our
 calculations reveal that both the interfaces with the
 non-magnetic metal V and the semiconductor InAs are no more
 half-metallic due to the different environment of the atoms of
 the half-metal at the interface. These interface states although
 localized only at the first few interface layers can become
 conducting when coupled to defect states and kill the
 spin-polarization of the current injected from the half-metal
 into the semiconductor or the non-magnetic metallic spacer.
\end{abstract}


\maketitle

\section{Introduction \label{sec1}}

The last decade one of the most active research areas in solid
state and materials science is the field of magnetoelectronics
also known as spintronics.\cite{Zutic,Felser,Zabel} The aim is to
replace conventional electronics by new devices where the spin of
the electrons and not the charge transfer plays the key role. A
central problem in this field is the injection of spin-polarized
current from a metal into a semiconductor.\cite{Wolf} In principle
it is possible to achieve 100\% spin-polarized injected current if
the magnetic lead is a half-metallic material. These alloys are
hybrids between normal metals and semiconductors. The Fermi level
crosses the majority spin-band as for a usual metal while it falls
within an energy gap in the minority spin-band as in a
semiconductor.\cite{Reviews,book} The attention on the half-metals
has been mainly concentrated on the intermetallic Heusler alloys
like NiMnSb (semi-Heuslers) or Co$_2$MnSn (full-Heuslers) due to
their very high Curie temperatures, which exceed considerably the
room temperature reaching even values close to 1000 K, and their
structural similarity to the widely used binary semiconductors
like GaAs.\cite{landolt1,landolt2} In such  compounds the behavior
of the interface between the half-metal and the semiconductor is
of great importance since interface states, although localized in
space, couple with impurity or defect states killing the
spin-polarization of the injected current.

In this communication we will concentrate on the properties of the
(001) surfaces of the half-metallic Heusler alloy Co$_2$MnSn and
its interfaces with the semiconductor InAs and the non-magnetic
metal V. Its bulk properties are well-established (see
\onlinecite{GalaFull} for example) but only the MnSn terminated
(001) surfaces have been theoretically studied using ab-initio
electronic structure calculations.\cite{Lee04} Theoretical
investigations exist for the (001) surfaces of Co$_2$MnSi and
Co$_2$MnGe compounds,\cite{GalaSurf} which are isovalence to
Co$_2$MnSn (same number of valence electrons in the unit cell) and
for the (001) surfaces of Co$_2$CrAl\cite{GalaSurf} and their
interfaces with the semiconductors GaAs\cite{Picozzi} and
InP\cite{GalaInter} respectively. In all cases the
half-metallicity was destroyed due to surfaces/interface states
which were strongly localized near the surface/interface layers.
Experimentally films of full-Heusler alloys have been grown by
several groups (see for example articles in reference
\onlinecite{Growth}) and they are widely used for realistic
applications like incorporated magnetic tunnel
junctions\cite{Marukame} and spin-valves.\cite{Kelekar}

We will start our study by presenting in section \ref{sec2} the
bulk properties of Co$_2$MnSn and the (001) surfaces. There are
two possible terminations : (i) a Co surface layer and a MnSn
subsurface layer and (ii) a MnSn surface layer and a pure Co
subsurface layers (for the exact structure see figure 1 in
reference \onlinecite{GalaInter}). In section \ref{sec3} we
present our results on the (001) interfaces with the InAs
semiconductor which has a lattice constant close to the one of
Co$_2$MnSn and thus pseudomorphic growth could be in principle
achieved and of the (001) interfaces with V since multilayers
between Heusler alloys and V have been succesfully grown
experimentally.\cite{Westerholt} InAs crystallizes in the
zinc-blende structure which is similar to the $L2_1$ structure of
the full-Heusler alloys (see reference \onlinecite{GalaInter}) and
V crystallizes in a b.c.c. lattice. The $L2_1$ structure of
full-Heusler alloys, if one ignores the different chemical
elements, is in reality a b.c.c. lattice and thus coherent growth
of V on top of Co$_2$MnSn is possible. In section \ref{sec4} we
summarize and conclude our results.

For the calculations we have used the the full-potential version
of the screened Korringa-Kohn-Rostoker (KKR) Green's function
method\cite{Pap02} in conjunction with the local spin-density
approximation\cite{vosko} for the exchange-correlation
potential.\cite{Kohn} The details for the surface and interfaces
calculations have been published
elsewhere.\cite{GalaSurf,GalaInter}

\begin{figure}
\includegraphics[width=\linewidth]{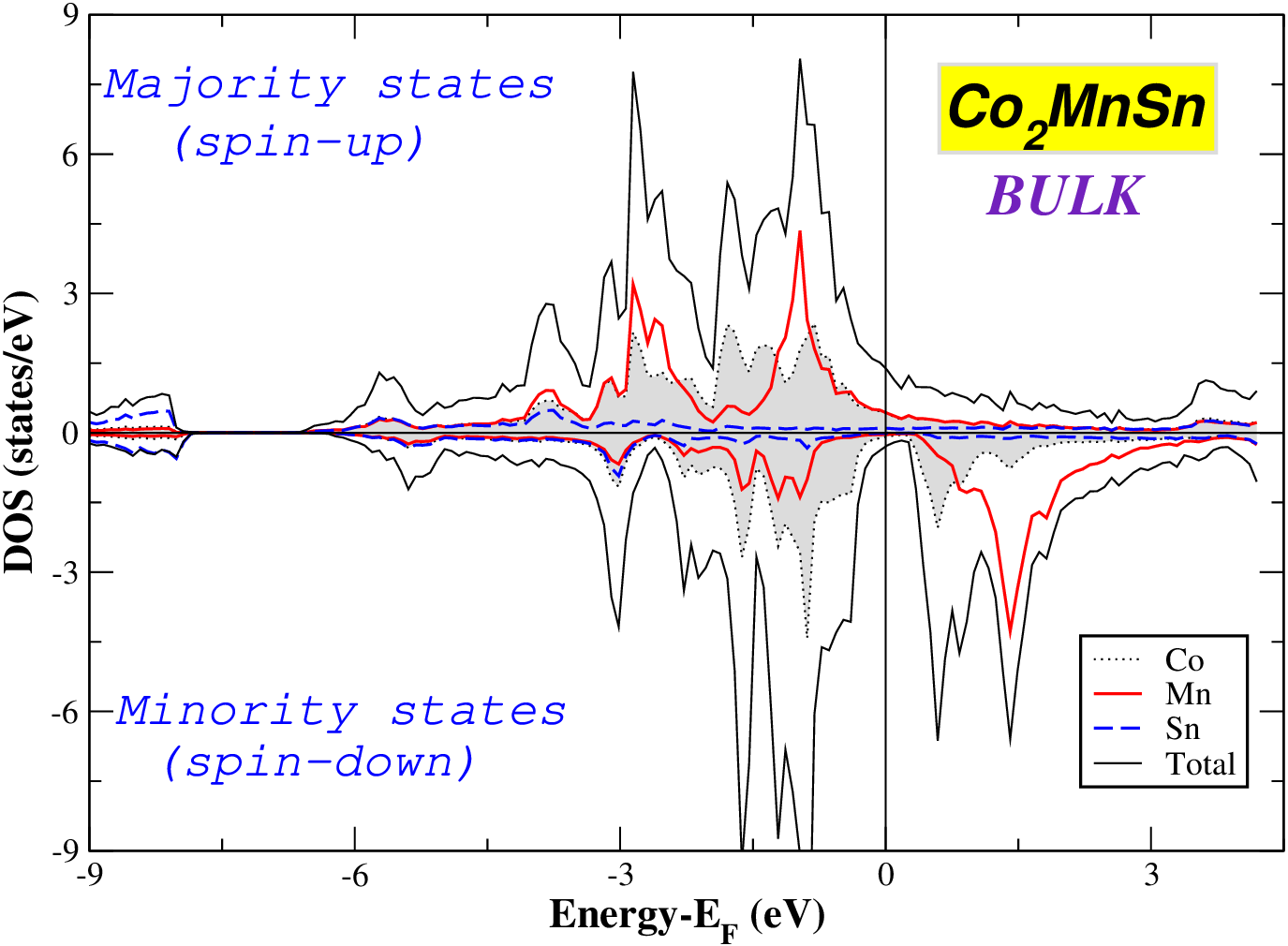}
\caption{(Color online) Total and atom-resolved density of states
(DOS) in states/eV as a function of the energy for the bulk
Co$_2$MnSn alloy. Positive values of DOS correspond to the
majority-spin (spin-up) electrons and negative values to the
minority (spin-down) electrons. We have chosen the zero energy to
correspond to the Fermi level.
 \label{fig1}}
\end{figure}

\section{Bulk and surface properties \label{sec2}}

First we will discuss the case of the bulk Co$_2$MnSn. In figure
\ref{fig1} we have plotted the density of states (DOS) in units
(number of states)/eV as a function of the energy in eV. The zero
of the energy axis corresponds to the Fermi level, $E_F$. We have
plotted both the total DOS and its projection on each atom. First,
the compound is half metal since the Fermi level crosses the
majority-spin (spin-up) electronic states while it is within a
tiny gap of the minority-spin (spin-down) states. At around -9 eV
are located the $s$ states provided by Sn. At around -6 eV start
the $p$ bands of the Sn atom which accommodate also $d$ electrons
of the transition metal atoms reducing the $d$-charge that should
be accommodated by their $d$ bands. At -3 eV start the $d$ bands
of the transition metal atoms. These states in the minority band
are mainly of Co character. As explained extensively in reference
\onlinecite{GalaFull} the gap is formed between states exclusively
located at the Co sites which due to symmetry reasons do not
hybridize with the $d$ orbitals of Mn. There are exactly 12
occupied minority states, one $s$ and three $p$ states of Sn, five
bonding $d$ Co-Mn states and finally the triple degenerated
$t_{1u}$ states which are exclusively located at Co sites as we
discussed before.\cite{GalaFull} Since there are exactly 12
occupied minority spin states the total spin moment should follow
the so-called Slater-Pauling behavior for the full-Heusler alloys
being in $\mu_B$ the total number of valence electrons in the unit
cell, which for Co$_2$MnSn is 29, minus two times the number of
occupied minority-spin states. Thus the total spin moment in the
unit cell should be 5 $\mu_B$. This value coincides with the
calculated total spin moment which is analyzed in table
\ref{table1}. Each Co atom has a spin moment of around 0.93
$\mu_B$ and the Mn atom has a large localized spin moment which is
3.20 $\mu_B$. The Sn atom has a small negative spin moment of
-0.06 $\mu_B$ since the bonding majority $p$ states of Sb are
spread over a wide energy range crossing the Fermi level while the
bonding minority $p$ states are completely occupied.

\begin{table}
\caption{Mn and Co atom-resolved spin magnetic moments in $\mu_B$.
For the surfaces and interfaces we present the spin-magnetic
moment of the atoms at the surface/interface layer and one layer
beneath.} \label{table1}
\begin{ruledtabular}
 \begin{tabular}{llcc}
 \multicolumn{2}{c}{Co$_2$MnSn}    & $m^{Co}$  &   $m^{Mn}$ \\
Bulk & & 0.93 & 3.20  \\
Surfaces & (001) Co-term & 1.80 & 3.35  \\
& (001) MnSn-term & 0.91 & 3.75  \\
Interfaces &Co/In & 2.00 & 3.19 \\
& Co/As & 1.35 & 3.42  \\
& Co/V & 0.92 & 3.51  \\
& MnSn/In & 1.82   & 2.73\\
& MnSn/As & 1.71  &  3.69 \\
& MnSn/V & 1.64 & 2.46  \\
\end{tabular}
\end{ruledtabular}
\end{table}

\begin{figure}
\includegraphics[width=\linewidth]{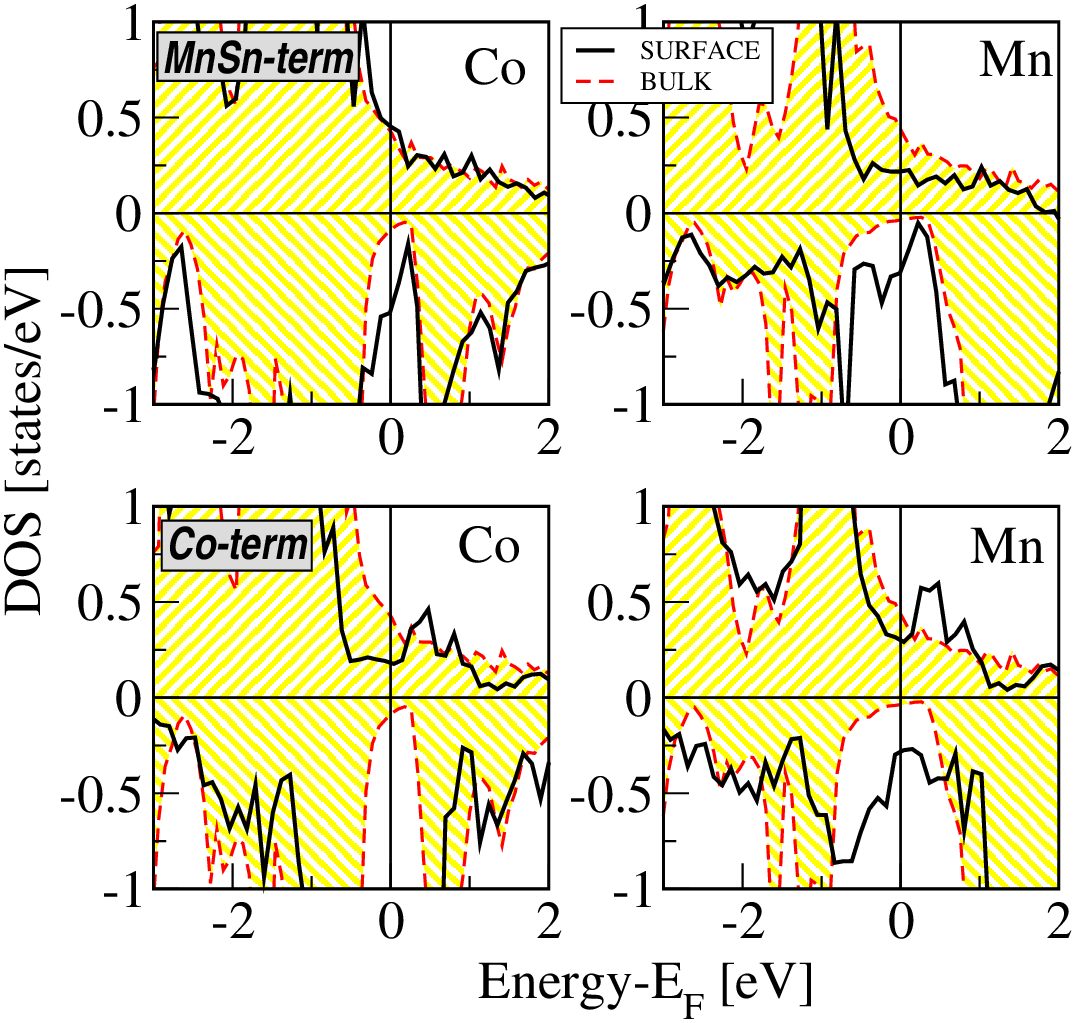}
\caption{(Color online) Upper panel : Atom-resolved DOS for the Mn
atom at the surface layer and the Co atom at the subsurface layer
in the case of the MnSn terminated Co$_2$MnSn(001) surface
compared to the bulk case. Lower panel : same for the Co
terminated Co$_2$MnSn(001) surface.
 \label{fig2}}
\end{figure}

In table \ref{table1} we have also included the spin moments of
the atoms at the surface and subsurface layers for both possible
Co and MnSn terminations and in figure \ref{fig2} we have included
the DOS of these atoms with respect to the bulk. When we open the
MnSn-terminated surface we break the bonds of the Mn and Sn atoms
with their nearest neighboring Co atoms (each Mn or Sn atom has
eight Co atoms as nearest neighbors and when we open the surface
we break four such bonds). Due to the creation of these ligand
states the Mn atoms regain some charge but since their majority
states are almost completely occupied and it will cost a lot in
Coulomb energy to occupy exclusively majority high-energy-lying
antibonding states, they have to accommodate the extra charge also
in minority-spin states . Thus although its spin moment only
slightly increases reaching the 3.75 $\mu_B$ half-metallicity is
completely destroyed since as can be seen in figure \ref{fig2}
there is a minority surface state pinned exactly at the Fermi
level which persists also for the Co atom at the subsurface layer.
The Co-terminated (001) surface presents even more strong
deviations from the perfect half-metallicity. Each Co atom at the
surface looses two out for its fours Mn neighbors and the symmetry
of its environment completely changes. As a result its spin
magnetic moment is almost doubled reaching the 1.80 $\mu_B$. The
unoccupied minority $d$ states are pushed lower in energy and the
Fermi level is now within a very large broad peak and
half-metallicity is completely destroyed. The altered Co DOS
polarizes also the DOS of the Mn atoms at the subsurface layer
which also shows an important minority DOS at the Fermi level. To
conclude the (001) surfaces of Co$_2$MnSn show no more the
half-metallic character of the perfect bulk compounds due to
surfaces states created by the ligands.

\section{Interfaces \label{sec3}}

\begin{figure}
\includegraphics[width=\linewidth]{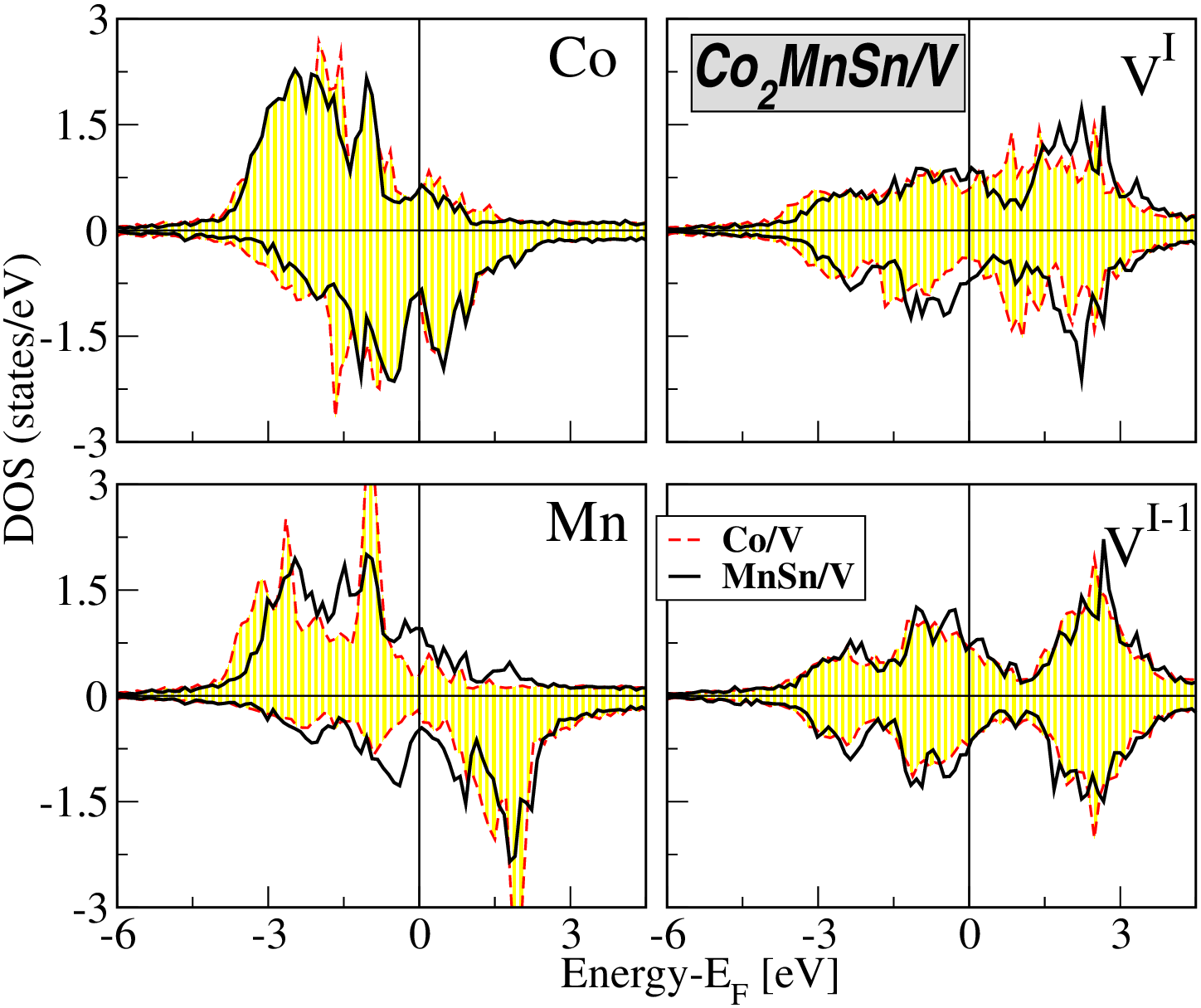}
\caption{(Color online) Atom-resolved DOS for the Co and Mn atoms
at the interface and subinterface layers and the V atom at the
interface (V$^\mathrm{I}$) and the subinterface (V$^\mathrm{I-1}$)
layer for both Co/V and MnSn/V (001) interfaces.
 \label{fig3}}
\end{figure}

We will start our discussion on the interfaces from the case of
interfaces with vanadium. Vanadium in its bulk form is non
magnetic but in proximity with strongly magnetic elements like Co
or Mn its $d$ orbitals can be polarized through hybridization
effects and it can obtain a spin magnetic moment. As in the case
of surfaces, the interface can be either made up from a V and a Co
interface layers or a V and a MnSn interface layers. In table
\ref{table1} we have included the spin magnetic moments of the Co
and Mn atoms at the interfaces since the Sn atoms present very
small magnetic moments and in figure \ref{fig3} we present the
DOS's of the atoms at the interface and subinterface layers for
both possible structures at the interface. When the interface is
made up by Co and V, the Co atoms at the interface have now from
one side two Mn atoms and two Sn atoms as nearest neighbors and
from the other side four V atoms. Thus their symmetry is very
close to the perfect $L2_1$ structure of the Heusler alloys and
they have a spin moment almost identical to the bulk case. Also
the spin moment of the Mn atoms at the subinterface layer is
slightly larger than the bulk case. But although the spin moments
for the Co/V case stay close to the bulk values, the DOS in figure
\ref{fig3} reveals that the half-metallicity is completely
destroyed since now the states around the minority gap strongly
overlap and the Fermi level falls within a deep of the DOS but not
a gap anymore. The V atoms at the interface, V$^\mathrm{I}$, are
polarized by the Co atoms carrying a small spin magnetic moment
while the V atoms at the subinterface layer, V$^\mathrm{I-1}$,
almost recover their bulk behavior and the DOS of the occupied
states is similar for both spin directions. In the case of the
MnSn/V interfaces the spin moments show larger deviations from the
ideal bulk case and half-metallicity is completely destroyed.

\begin{figure*}
\includegraphics[scale=0.45]{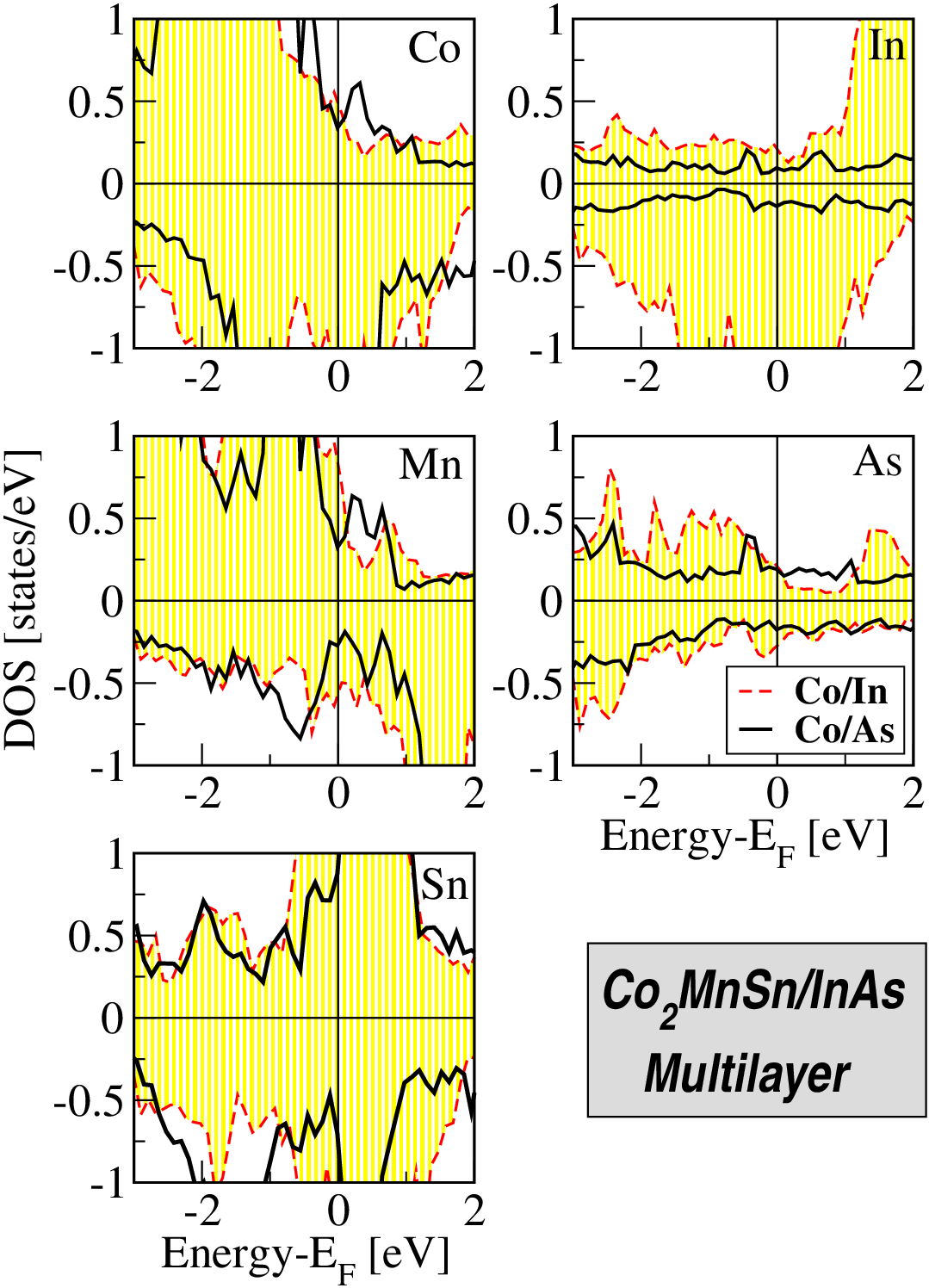}
\includegraphics[scale=0.45]{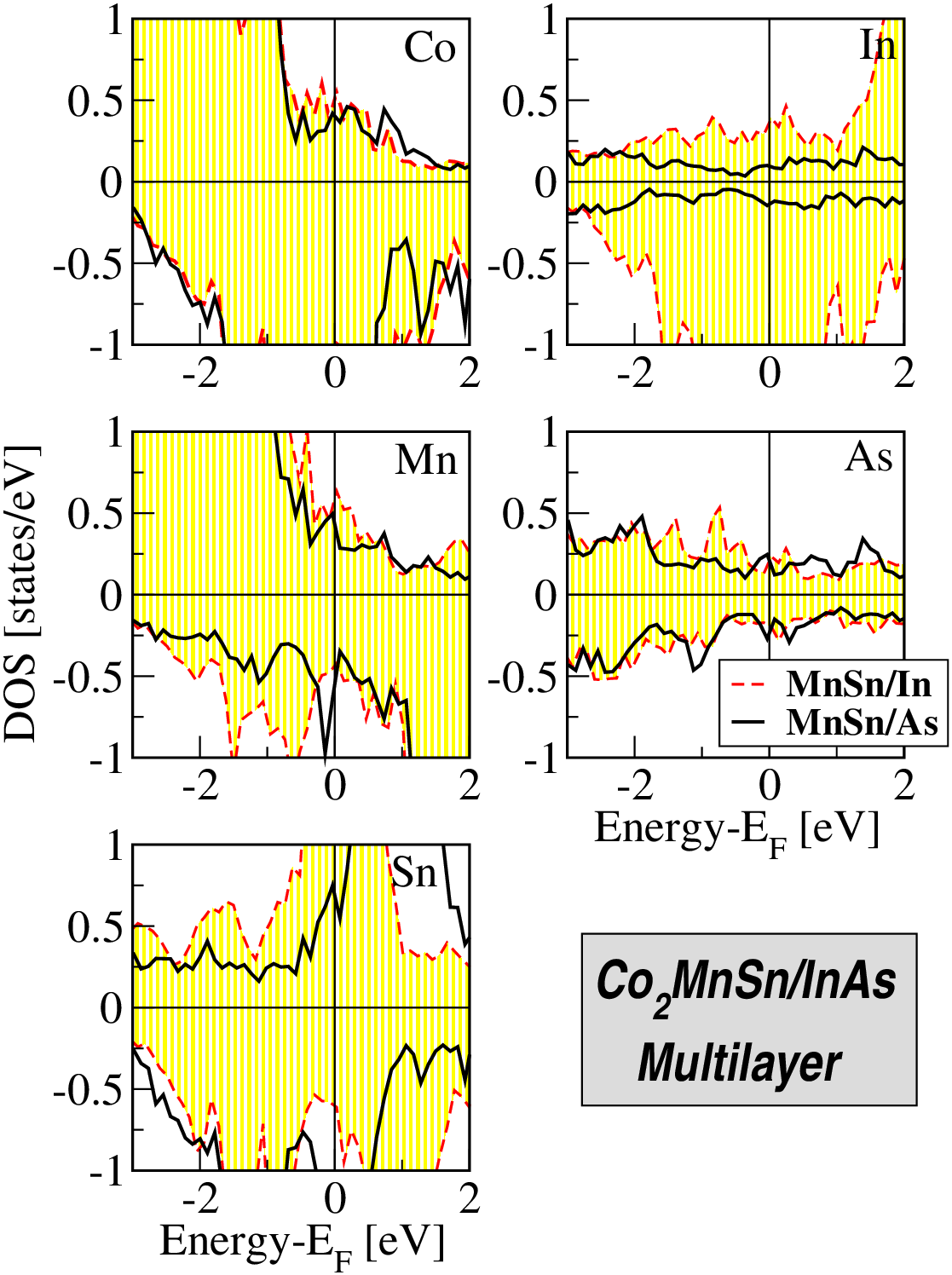}
\caption{(Color online) Left panel : Atom-resolved DOS for the Co
and In interface atoms in the case of the Co/In interface (or Co
and As in the case of the Co/As interface)  and Mn,Sn and As atom
at the sub-interface layers in the case of the Co/In interface (or
Mn, Sn and In in the case of the Co/As interface). Right panel :
same for the MnSn/In and MnSn/As interfaces.
 \label{fig4}}
\end{figure*}

We will complete our study by the case of interfaces with the InAs
semiconductor. There are four different possible interfaces Co/In,
Co/As, MnSn/In and MnSn/As. A close look at table \ref{table1}
reveals that for all four interfaces Co at the interface layer or
subinterface layers has a spin moment which is from 50\%\ larger
than the bulk value and reaches a value of 2 $\mu_B$ at the Co/In
interface. In the case of the Co/As interface the spin moment is
smaller than this maximum value (1.35 $\mu_B$) since the Co atoms
at the interface have from the side of the half-metal two Mn and
two Sn atoms as first neighbors and from the side of the
semiconductor only two As atoms. Thus at the Co/As interface we
have broken two out for the four Co-Mn bonds. Indium has a
different number of valence electrons with respect to As and Sn
and the environment of the Co atoms at the Co/In interface is
considerably altered with respect to the bulk and this leads to
this huge spin moment of 2 $\mu_B$  for the Co interface atoms. In
the case of MnSn/In and MnSn/As interfaces the interface deviates
even more from the bulk case since the Mn and Sn atoms at the
interface have now from the side of the semiconductor two In or
two As atoms as first neighbors, respectively, instead of the two
Co atoms in the bulk. This is reflected on the large fluctuations
of the Mn spin moments and the large spin magnetic moments of the
Co atoms at the subinterface layers. As can be seen in figure
\ref{fig4} the half-metallicity is in all cases destroyed since
even the In and As atoms of the semiconductor at the interface
present DOS around the Fermi level due to the hybridization of
their $p$ orbitals with the $p$ and $t_{2g}$ orbitals of the Co or
Mn-Sn atoms at the interface layers of the Co$_2$MnSn alloy.

\section{Summary and conclusions \label{sec4}}

We have performed state-of-the-art electronic structure
calculations for the Co$_2$MnSn full-Heusler alloy. Our results
for the bulk show that the alloy in form of single crystals is a
half-metallic ferromagnet with the Fermi level being located
within a tiny gap of the minority-spin DOS. The alloy shows the
Slater-Pauling behavior with a total spin magnetic moment in the
unit cell of 5 $\mu_B$. In the case of the MnSn terminated
 (001) surface, the broken bonds at the surface form a minority band
 pinned exactly at the Fermi level, while for the Co-terminated
 (001) surface the effect is even more pronounced. Our
 calculations reveal that both the interfaces with the
 non-magnetic metal V and the semiconductor InAs are no more
 half-metallic due to the different environment of the atoms of
 the half-metal at the interface. These interface states although
 localized only at the first few interface layers can become
 conducting when coupled to defect states and kill the
 spin-polarization of the current injected from the half-metal
 into the semiconductor.

%---------------------------------------------------------------------------------------------------------------------------

\end{document}